\documentclass[11pt]{article}
\usepackage[T1]{fontenc}
\usepackage[latin9]{inputenc}
\usepackage{float,url,amsmath,amssymb}
\usepackage{mathrsfs}   
\usepackage[english]{babel}
\usepackage[absolute]{textpos}

\usepackage{multirow}
\begin{document}

\title{Calculating the renormalisation group equations of a SUSY model with
\texttt{Susyno}}

\author{Renato M. Fonseca \\ \\ Centro de Física Teórica de Partículas,
Instituto Superior Técnico,\\
Universidade Técnica de Lisboa, Av. Rovisco Pais 1, 1049-001 Lisboa,
Portugal \\ \\ \textit{E-mail address:} renato.fonseca@ist.utl.pt}
\date{}

\maketitle

\setlength{\TPHorizModule}{\paperwidth}\setlength{\TPVertModule}{\paperheight}
\begin{textblock}{0.2}(0.8,0.08)
\begin{center}
CFTP/11-011\\
\today
\end{center}
\end{textblock}

\begin{abstract}
\texttt{Susyno} is a Mathematica package dedicated to the computation of
the 2-loop renormalisation group equations of a supersymmetric model based on
any gauge group (the only exception being multiple U(1) groups) and for any
field content.\end{abstract}

\section{Introduction}

Supersymmetry (SUSY) is an elegant and well known extension of the
Standard Model (SM) which aims at solving or softening some of the
SM theoretical and experimental shortcomings. Supersymmetric models
based on Grand Unified Theories (GUTs) offer the appealing
feature of being described by a unique gauge group and coupling
constant; furthermore, the potential new superfields, as well as  their dynamics,
can provide answers to fundamental problems. For example SO(10) based
models offer a natural explanation for neutrino masses and
mixings via a type-I seesaw. This occurs since the spinor representation
of the group contains a SM singlet superfield (with the appropriate
quantum numbers). Even larger and/or more complex groups are often considered, for
example $E_{6}$ in string inspired models.

The analysis of the theoretical and phenomenological implications
of SUSY GUT models requires a careful study of the evolution of the
fundamental parameters from the high-energy scale down to the electroweak
scale, at which observables are computed and constraints applied.
As such, knowledge of the renormalisation group equations (RGEs) is
necessary. Although the RGEs of several models (e.g., MSSM, NMSSM)
are already known~\cite{MartinVaughn:RGEs,Ellwanger:2009dp}, for
other SUSY extensions of the SM complicated general equations must
be used~\cite{MartinVaughn:RGEs,Yamada:1994id}.

Here we describe \texttt{Susyno}, a Mathematica-based package that addresses
this issue. The program takes as input the gauge group, the representations
(i.e., the chiral superfield content), the number of flavours/copies%
\footnote{Throughout this document, the word \textit{flavours} is
  meant to convey the sense of \textit{copies} or \textit{repetitions}
  of a given representation/field. %
} of each representation, and any abelian discrete symmetries (e.g.,
R-parity). \texttt{Susyno} then constructs
the most general superpotential and soft-SUSY-breaking Lagrangian 
consistent with the field content and symmetries
imposed. Once these elements have been derived, 
\texttt{Susyno} then computes the 2-loop
$\beta$-functions of all the parameters of the model, which is its main output.

There is another Mathematica package, \texttt{SARAH}~\cite{Staub:2008uz},
which works in a similar way for models based on SU($n$) gauge factor groups.
However, \texttt{Susyno} is prepared to accept as working input any gauge group that
does not contain more that one U(1) factor\footnote{Multiple U(1) groups lead
 to ``U(1) mixing''~\cite{U1_Mixing} which requires special care~\cite{U1_MixingRGEs}.%
}.

This document is organised as follows. Section~\ref{sec:Installation-and-quick}
explains how to install the program and run a first simple example
(the MSSM). Sections~\ref{sec:Input} and~\ref{sec:Output} explain
how to prepare the input and how to read and interpret the output, also using
as example the MSSM case. 
Finally, Section~\ref{sec:Tests-made} summarises the tests conducted
to validate the code.

\section{\label{sec:Installation-and-quick}Installation and quick start}

\texttt{Susyno} works on Windows, Linux and Mac OS provided that Mathematica
7 (or a latter version) is installed. 
The program is obtainable from\\

\noindent \url{http://web.ist.utl.pt/renato.fonseca/susyno.html}\\

The files 
\textit{LieGroups.m}, \textit{SusyRGEs.m} and \textit{Susyno.m} are
the core of the program. These and other auxiliary files can be found inside
the folder \textit{Susyno}, which must be extracted from the downloaded
\textit{Susyno-1.1.tar} file to a location that is visible to Mathematica. 
Typing \texttt{\$Path} in Mathematica will show a complete list of acceptable locations.
A good choice is to place the whole folder (not just its contents) in\\

\noindent \textit{(Mathematica base directory)/AddOns/Applications}\\

\noindent (note that in a Windows system the slashes ``/'' 
must be replaced by backslashes ``\textbackslash''). The package can be loaded by typing\\

\noindent \texttt{<\textcompwordmark{}< Susyno$\grave{\,\textrm{\,}}$}~\\

\noindent in Mathematica's front end. A text message is returned,
informing that a built-in help system provides a detailed
description of the program and its functions (see \ref{sec:List-of-available}).
A tutorial is also included.

The \texttt{Susyno} lines below allow a simple and easy first run: 
the example consists of a possible way of writing the MSSM input. \\

\noindent \texttt{normalisation = Sqrt{[}3/5{]};}

\noindent \texttt{Q = \{U1->1/6 normalisation, SU2->\{1\}, SU3->\{1, 0\}, NFlavours->3, DiscreteSym->-1\};}

\noindent \texttt{u = \{U1->-2/3 normalisation, SU2->\{0\}, SU3->\{0, 1\}, NFlavours->3, DiscreteSym->-1\};}

\noindent \texttt{d = \{U1->1/3 normalisation, SU2->\{0\}, SU3->\{0, 1\}, NFlavours->3, DiscreteSym->-1\};}

\noindent \texttt{L = \{U1->-1/2 normalisation, SU2->\{1\}, SU3->\{0, 0\}, NFlavours->3, DiscreteSym->-1\};}

\noindent \texttt{e = \{U1->normalisation, SU2->\{0\}, SU3->\{0, 0\}, NFlavours->3, DiscreteSym->-1\};}

\noindent \texttt{Hu = \{U1->1/2 normalisation, SU2->\{1\}, SU3->\{0, 0\}, NFlavours->1, DiscreteSym->1\};}

\noindent \texttt{Hd = \{U1->-1/2 normalisation, SU2->\{1\}, SU3->\{0, 0\}, NFlavours->1, DiscreteSym->1\};}~\\

\noindent \texttt{model = \{Q, u, d, L, e, Hu, Hd\};}

\noindent \texttt{BetaFunctions2L{[}model{]};}~\\

\noindent 
Evaluation of this simple code generates the 2-loop $\beta$-functions
of the model (MSSM in this case).
Notice that no external input or output files are used - everything happens on Mathematica's
front end.

\section{\label{sec:Input}Defining a model}
A SUSY model contains two building blocks: a superpotential and a soft-SUSY-breaking
Lagrangian. For a simple gauge group, a general superpotential can be written as 
\begin{equation}
W = 
\frac{1}{6}Y^{ijk}\Phi_{i}\Phi_{j}\Phi_{k}+\frac{1}{2}\mu^{ij}\Phi_{i}\Phi_{j}+L^{i}\Phi_{i}\,,
\end{equation}
with $\Phi$ denoting a chiral superfield, and where $Y$, $\mu$ and $L$
are dimensionless, mass, and mass$^2$ parameters, respectively.
A generic soft-SUSY-breaking Lagrangian reads
\begin{equation}
-\mathscr{L}_{soft} = 
\left(\frac{1}{6}h^{ijk}\phi_{i}\phi_{j}\phi_{k}+
\frac{1}{2}b^{ij}\phi_{i}\phi_{j}+s^{i}\phi_{i}+\textrm{h.c.}\right)+
\left(m^{2}\right)_{j}^{i}\phi_{i}\phi_{j}^{*}+\left(\frac{1}{2}M\lambda\lambda+
\textrm{h.c.}\right)
\end{equation}
where $\lambda$ is a gaugino field and $\phi$ denotes scalar
fields. $M$ ($m^2$) correspond to gaugino (scalar) soft breaking
masses, while $h$, $b$ and $s$ denote trilinear, bilinear and linear
soft breaking parameters. \\
\texttt{Susyno} requires as input the list of chiral superfields (just
\textit{fields} from now on) of a model:\\

\noindent \texttt{model=\{field1, field2,...\};}~\\

\noindent For each field, the user only needs to provide its defining
elements. The syntax is\\

\noindent \texttt{field=\{g1->g1\_rep, g2->g2\_rep, ..., NFlavours->nf,
DiscreteSym->q\};}~\\

\noindent where the values to be filled are the following:
\begin{itemize}
\item \texttt{g1}, \texttt{g2}, ... (gauge factor groups);
\item \texttt{g1\_rep}, \texttt{g2\_rep}, ... (representations of the field
under each of the gauge factor groups);
\item \texttt{nf} (number of flavours);
\item \texttt{q} (charge under some discrete abelian symmetry).
\end{itemize}
To obtain the $\beta$-functions, the user must call \texttt{BetaFunctions1L}
(or \texttt{BetaFunctions2L} for 2-loop results):\\

\noindent \texttt{BetaFunctions1L{[}model{]}}

\noindent \texttt{BetaFunctions2L{[}model{]}}~\\

\noindent or\\

\noindent \texttt{BetaFunctions1L{[}model,Verbose->False{]};}

\noindent \texttt{BetaFunctions2L{[}model,Verbose->False{]};}~\\

\noindent if we do not wish to have the results printed on Mathematica's
front end.

With this information the program then builds internally the superpotential
and the soft-SUSY-breaking Lagrangian using an algorithm to automatically
name the parameters of the model (see Section 4). Assuming that the
model contains parameters named \texttt{X1}, \texttt{X2}, ..., the
program's output is\\

\noindent \texttt{\{\{X1, $\beta_{\mathtt{X1}}^{\left(1\right)}$\},\{X2,
$\beta_{\mathtt{X2}}^{\left(1\right)}$\},...\}}~~~~~~~~~~~~~~~~~~~~~~~~~~~~~~~~~~~~~~~~~~~~~~~~~~~(\texttt{BetaFunctions1L})

\noindent \texttt{\{\{X1, $\beta_{\mathtt{X1}}^{\left(1\right)}$,
$\beta_{\mathtt{X1}}^{\left(2\right)}$\},\{X2, $\beta_{\mathtt{X2}}^{\left(1\right)}$,
$\beta_{\mathtt{X2}}^{\left(2\right)}$\},...\}}~~~~~~~~~~~~~~~~~~~~~~~~~~~~~~~~~~(\texttt{BetaFunctions2L})\\

\noindent that is, a set of sublists each containing a parameter \texttt{Xi},
and the one-loop (one- and two-loop) $\beta$-functions for its RG
evolution: $\beta_{\mathtt{Xi}}^{\left(1\right)}$ ($\beta_{\mathtt{Xi}}^{\left(1\right)}$,
$\beta_{\mathtt{Xi}}^{\left(2\right)}$).

In the following, we will look at each of the input elements mentioned
above in greater detail.

\subsection{Gauge factor groups}

The program needs to know the model's gauge group%
\footnote{We emphasise here that we are actually dealing with algebras, not
groups. Nevertheless, we will adopt the common practice in high-energy
physics of using the word \textit{group} for both these concepts.%
}, which it extracts from each \texttt{field} object mentioned above.
Therefore the gauge factor groups should be presented in the same
order in all fields. In addition, U(1)'s must always come first. For
the MSSM (for completeness, the superpotential and soft-SUSY-breaking
Lagrangian of the MSSM are presented in~\ref{sec:Parameters-of-the-MSSM})
we could write each field as \\

\noindent \texttt{field = \{U1->U1\_charge,SU2->SU2\_rep,SU3->SU3\_rep,...\};}~\\

\noindent We may change the order of SU(2) and SU(3) but the U(1)
factor must always come first.

\noindent Any group can be used: all the possible gauge factor groups,
as well as their corresponding \texttt{Susyno} input are collected
in Table~\ref{table:gaugefactor}. 
\begin{table}[tbph]
\begin{centering}
\begin{tabular}{cc}
Gauge factor group  & \texttt{Susyno} input\tabularnewline
\hline 
U(1)  & \texttt{U1}\tabularnewline
SU($n$)  & \texttt{SU2}, \texttt{SU3}, \texttt{SU4}, \texttt{SU5}, ...\tabularnewline
SO($n$)  & \texttt{SO3}, \texttt{SO5}, \texttt{SO6}, \texttt{SO7}, ...\tabularnewline
Sp($2n$)  & \texttt{SP2}, \texttt{SP4}, \texttt{SP6}, ...\tabularnewline
$\mathrm{G}_{2}$  & \texttt{G2}\tabularnewline
$\mathrm{F}_{4}$  & \texttt{F4}\tabularnewline
$\mathrm{E}_{6}$, $\mathrm{E}_{7}$, $\mathrm{E}_{8}$  & \texttt{E6}, \texttt{E7}, \texttt{E8}\tabularnewline
\end{tabular}
\par\end{centering}

\caption{Gauge factor groups}

\label{table:gaugefactor} 
\end{table}

Keep in mind that despite its generality, \texttt{Susyno} only assigns to
variables a finite number of groups (for instance variable \texttt{SO100}
is not set). Should this ever become a problem, there is an easy method
of circumventing it, discussed in~\ref{sec:More-on-groups-reps}.

\subsection{Representations}

As mentioned before, \texttt{Susyno} is designed to accept an arbitrary
field content so any representation of any gauge group is allowed.
For U(1) factor groups, all we need are the hypercharges of each field,
which are just numbers%
\footnote{Notice that, while it is possible to have more than one hypercharge
(i.e. more than one U(1) factor group), the formalism of \texttt{Susyno}
is based on~\cite{MartinVaughn:RGEs,Yamada:1994id}, which do not
include the effects of a possible U(1) mixing.%
}. On the other hand, the representations of simple gauge groups must
be specified by their Dynkin coefficients (see~\ref{sec:More-on-groups-reps}
for details). In Table~\ref{tab:Table_representations} we list some
of the representations of SU(2), SU(3), SU(5) and SO(10) (Dynkin coefficients
and corresponding dimensions). 
\begin{table}[h!]
\begin{centering}
\begin{tabular}{lrl}
\textbf{Group}  & \multicolumn{2}{c}{\textbf{Representation}}\tabularnewline
\hline 
 & \textit{Dynkin coefficients}  & \textit{Dimension (Name)}\tabularnewline
SU(2)  & \{0\}  & $\boldsymbol{1}$ (Trivial/Singlet)\tabularnewline
 & \{1\}  & $\boldsymbol{2}$ (Fundamental/Doublet)\tabularnewline
 & \{2\}  & $\boldsymbol{3}$ (Adjoint/Triplet)\tabularnewline
SU(3)  & \{0,0\}  & $\boldsymbol{1}$ (Trivial/Singlet)\tabularnewline
 & \{1,0\}  & $\boldsymbol{3}$ (Fundamental)\tabularnewline
 & \{0,1\}  & $\overline{\boldsymbol{3}}$ (Anti-fundamental)\tabularnewline
 & \{1,1\}  & $\boldsymbol{8}$ (Adjoint)\tabularnewline
SU(5)  & \{0,0,0,0\}  & $\boldsymbol{1}$ (Trivial/Singlet)\tabularnewline
 & \{1,0,0,0\}  & $\boldsymbol{5}$ (Fundamental)\tabularnewline
 & \{0,0,0,1\}  & $\overline{\boldsymbol{5}}$ (Anti-fundamental)\tabularnewline
 & \{0,1,0,0\}  & $\boldsymbol{10}$\tabularnewline
 & \{2,0,0,0\}  & $\boldsymbol{15}$\tabularnewline
 & \{0,0,0,2\}  & $\overline{\boldsymbol{15}}$\tabularnewline
 & \{1,0,0,1\}  & $\boldsymbol{24}$ (Adjoint)\tabularnewline
SO(10)  & \{0,0,0,0,0\}  & $\boldsymbol{1}$ (Trivial/Singlet)\tabularnewline
 & \{1,0,0,0,0\}  & $\boldsymbol{10}$ (Fundamental)\tabularnewline
 & \{0,0,0,0,1\}  & $\boldsymbol{16}$ (Spinor)\tabularnewline
 & \{0,0,0,1,0\}  & $\overline{\boldsymbol{16}}$ (Spinor's conjugate)\tabularnewline
 & \{0,1,0,0,0\}  & $\boldsymbol{45}$ (Adjoint)\tabularnewline
 & \{2,0,0,0,0\}  & $\boldsymbol{54}$\tabularnewline
 & \{0,0,1,0,0\}  & $\boldsymbol{120}$\tabularnewline
 & \{0,0,0,0,2\}  & $\boldsymbol{126}$\tabularnewline
 & \{0,0,0,2,0\}  & $\overline{\boldsymbol{126}}$\tabularnewline
\end{tabular}
\par\end{centering}

\caption{\label{tab:Table_representations}List of some of the representations
of SU(2), SU(3), SU(5) and SO(10)}
\end{table}

To understand how the MSSM was specified in the example of Section~\ref{sec:Installation-and-quick},
we just need the following information from Table~\ref{tab:Table_representations}: 
\begin{itemize}
\item the Dynkin coefficients of the trivial and fundamental representations
of SU(2): \{0\} and \{1\}; 
\item the Dynkin coefficients of the trivial, fundamental ($\boldsymbol{3}$)
and anti-fundamental ($\overline{\boldsymbol{3}}$) representations
of SU(3): \{0,0\}, \{1,0\}, \{0,1\}. 
\end{itemize}
\noindent Hence\\

\noindent \texttt{Q = \{U1->1/6 normalisation, SU2->\{1\}, SU3->\{1,
0\}, ...\};}

\noindent \texttt{u = \{U1->-2/3 normalisation, SU2->\{0\}, SU3->\{0,
1\}, ...\};}

\noindent \texttt{d = \{U1->1/3 normalisation, SU2->\{0\}, SU3->\{0,
1\}, ...\};}

\noindent \texttt{L = \{U1->-1/2 normalisation, SU2->\{1\}, SU3->\{0,
0\}, ...\};}

\noindent \texttt{e = \{U1->normalisation, SU2->\{0\}, SU3->\{0, 0\},
...\};}

\noindent \texttt{Hu = \{U1->1/2 normalisation, SU2->\{1\}, SU3->\{0,
0\}, ...\};}

\noindent \texttt{Hd = \{U1->-1/2 normalisation, SU2->\{1\}, SU3->\{0,
0\}, ...\};}~\\

\noindent Notice that \texttt{Susyno} accepts any normalisation of
the hypercharges. In the sample code of Section~\ref{sec:Installation-and-quick},
we used the usual $\sqrt{\frac{3}{5}}$ factor (from an embedding
of the MSSM in an SU(5) based model).

\subsection{Number of flavours and abelian discrete symmetries}

Many models contain repetitions of some of the representations of
the gauge group. Instead of considering them distinct fields, we usually
view these as different flavours of a single field and \texttt{Susyno}
needs to know how many flavours there are in order to compute the
$\beta$-functions. The user should use the syntax \texttt{NFlavours->...}
for this purpose. For example, in the MSSM we have\\
\\
\texttt{Q = \{..., NFlavours->3, ...\};}

\noindent \texttt{u = \{..., NFlavours->3, ...\};}

\noindent \texttt{d = \{..., NFlavours->3, ...\};}

\noindent \texttt{L = \{..., NFlavours->3, ...\};}

\noindent \texttt{e = \{..., NFlavours->3, ...\};}

\noindent \texttt{Hu = \{..., NFlavours->1, ...\};}

\noindent \texttt{Hd = \{..., NFlavours->1, ...\};}~\\

Let us consider another example: if we were to modify the MSSM to
include $m$ copies of $H_{u}$ and $H_{d}$, we would write\\

\noindent \texttt{Hu = \{..., NFlavours->m, ...\};}

\noindent \texttt{Hd = \{..., NFlavours->m, ...\};}~\\

Finally, there are models in which we must forbid some couplings and
in order to do that some abelian discrete symmetry is introduced.
These symmetries are also a defining element of these models and as
such each field's charge must be passed on to the program using the
expression \texttt{DiscreteSym->...}. The MSSM is one such case;
there is a $Z_{2}$ symmetry known as R-parity so we should write\\
\\
\texttt{Q = \{..., DiscreteSym->-1\};}

\noindent \texttt{u = \{..., DiscreteSym->-1\};}

\noindent \texttt{d = \{..., DiscreteSym->-1\};}

\noindent \texttt{L = \{..., DiscreteSym->-1\};}

\noindent \texttt{e = \{..., DiscreteSym->-1\};}

\noindent \texttt{Hu = \{..., DiscreteSym->1\};}

\noindent \texttt{Hd = \{..., DiscreteSym->1\};}~\\

\noindent For the most general (R-parity violating) MSSM the first
5 lines must be replaced by\\

\noindent \texttt{Q = \{..., DiscreteSym->1\};}

\noindent \texttt{u = \{..., DiscreteSym->1\};}

\noindent \texttt{d = \{..., DiscreteSym->1\};}

\noindent \texttt{L = \{..., DiscreteSym->1\};}

\noindent \texttt{e = \{..., DiscreteSym->1\};}~\\

Note that the expressions \texttt{NFlavours->...} and \texttt{DiscreteSym->...}
can be placed anywhere along the definition of each field. They can
even be omitted: if no \texttt{NFlavours} is written, the program
will use the symbolic value \texttt{nf{[}i{]}} where $i$ is the order
by which a field is given; if there is no \texttt{DiscreteSym} then
\texttt{Susyno} will consider that the field transforms trivially
(charge equal to 1).

\section{\label{sec:Output}The output of \texttt{Susyno}}

Once all the input is provided, \texttt{Susyno} automatically 
builds the Lagrangian of a model, asking neither
for names of fields nor names of parameters. On the one hand, 
inputting a model becomes very easy, since it is not even necessary to know
the exact number of its parameters. On the other hand, this renders
the output hard to read (and hence not particularly user-friendly), 
since the names of the  parameters are chosen by the program.

Still, provided we know the conventions used by the program, 
the notation used by \texttt{Susyno} can be manually changed by the
user (after the code is executed). 
We notice that in the present version of \texttt{Susyno} 
there are no custom built-in functions to export the
results. Users who wish to do so must do it manually or with the help
of Mathematica's built-in functions \texttt{CForm} and \texttt{FortranForm}.

\subsection{Naming of parameters}

\texttt{Susyno} assigns names to the parameters of a model in such a
way that the user can identify which fields they are multiplying:\\

\texttt{y{[}field1, field2, field3, InvIndex, <flav1>, <flav2>, <flav3>{]}}

\texttt{mu{[}field1, field2, InvIndex, <flav1>, <flav2>{]}}

\texttt{l{[}field1, InvIndex, <flav1>{]}}

\texttt{h{[}field1, field2, field3, InvIndex, <flav1>, <flav2>, <flav3>{]}}

\texttt{b{[}field1, field2, InvIndex, <flav1>, <flav2>{]}}

\texttt{s{[}field1, InvIndex, <flav1>{]}}

\texttt{m2{[}field1, field2, InvIndex, <flav1>, <flav2>{]}}\\

A few comments concerning the above (output) tensors are in order:
\begin{itemize}
\item \texttt{y}, \texttt{mu}, \texttt{l}, \texttt{h}, \texttt{b}, \texttt{s}
and \texttt{m2} can be easily identified with the different types of
couplings and dimensionful parameters
of the superpotential and the soft-SUSY-breaking Lagrangian;

\item \texttt{field1}, \texttt{field2}, \texttt{field3} are the indices
of the fields entering a given coupling, or soft breaking term. 
For example, with model=\{Q,
u, d, L, e, Hu, Hd\} we have Q=1, u=2, d=3, L=4, e=5, Hu=6, Hd=7.
The up-quark Yukawa couplings would then be \texttt{y{[}1,2,6,...{]};}

\item There is the possibility that the product of 3 representations, 
$R_{1}\otimes R_{2}\otimes R_{3}$, contains more than one invariant. 
Therefore we need \texttt{InvIndex=1,2,...} to
distinguish them. This is rare though, so in most cases (e.g., the
MSSM) \texttt{InvIndex=1} for all parameters. Notice that in linear
and bilinear terms, $R_{1}$ and $R_{1}\otimes R_{2}$, this problem
does not arise since there is at most one invariant. However, for
consistency \texttt{Susyno} still uses \texttt{InvIndex} in this case,
although setting its value to \texttt{1};

\item \texttt{<flav1>}, \texttt{<flav2>} , \texttt{<flav3>} are the flavour
indices of \texttt{field1}, \texttt{field2}, \texttt{field3}. If any
of these fields has only one flavour, the corresponding index is omitted.
Consider again the example of the up-quark Yukawa couplings: 
we would have \texttt{y{[}1,2,6,1,i,j,k{]}}
where \texttt{i} = flavour of $Q$, \texttt{j} = flavour of u, \texttt{k}
= flavour of Hu. Yet Hu only has one flavour so the correct parameter
name is \texttt{y{[}1,2,6,1,i,j{]}}.
\end{itemize}

Additionally, there are also the coupling constants and the gaugino masses:\\

\noindent \texttt{g{[}1{]}, g{[}2{]}, ...}

\noindent \texttt{M{[}1{]}, M{[}2{]}, ...}\\

A list containing the matching of
MSSM's standard notation parameters and \texttt{Susyno} assignements
can be found in~\ref{sec:Parameters-of-the-MSSM}.

\subsection{The full Lagrangian as used by the program}

Knowing the full Lagrangian can be important. As an example, consider
a possible normalisation issue: how can we be sure that \texttt{Susyno}
is correctly taking $\mu H_{u}\cdot H_{d}$ as \texttt{mu{[}6,7,1{]}}$H_{u}\cdot H_{d}$,
and not -\texttt{mu{[}6,7,1{]}}$H_{u}\cdot H_{d}$, 2\texttt{mu{[}6,7,1{]}}$H_{u}\cdot H_{d}$
or any other multiple? The way to solve such problems related to the
normalisation of the parameters is to check the Lagrangian. This can
be done using the \texttt{ShowLagrangian} function:\\

\noindent \texttt{ShowLagrangian{[}model{]}}~\\

This function returns the list of tensors $\left\{ Y,\mu,L,h,b,s,m^{2}\right\} $.
Note that \texttt{Susyno} never expands the flavour indices of these
tensors so the first, second and third indices are assumed to carry
symbolic flavours $i$ (or none), $j$ (or none), $k$ (or none)%
\footnote{These flavour indices do not show up if the particular entry of the
tensors we are considering is unflavoured. For example the entry of
the Y tensor related to the fields Q, d, Hd will depend on $i$ (flavour
of the first field, Q) and $j$ (flavour of the second field, d) but
not on $k$ since the third field (Hd) is unflavoured. Notice also
that exchanging for example Hd with Q corresponds to having a different
entry of the Y tensor: this is because the flavour of Q (now the third
field) is assumed to be symbolically $k$ instead of $i$. %
}. The argument (\texttt{model}) is the same as in the \texttt{BetaFunctions1L}
and \texttt{BetaFunctions2L} functions. There is also a \texttt{Verbose}
option (set to \texttt{True} by default) that prints the different
parts of $W$ and $\mathscr{L}_{soft}$: 
\begin{itemize}
\item $\frac{1}{6}Y^{ijk}\Phi_{i}\Phi_{j}\Phi_{k}$ (Y part), $\frac{1}{2}\mu^{ij}\Phi_{i}\Phi_{j}$
($\mu$ part), $L^{i}\Phi_{i}$ (L part); 
\item $\frac{1}{6}h^{ijk}\phi_{i}\phi_{j}\phi_{k}$ (H part), $\frac{1}{2}b^{ij}\phi_{i}\phi_{j}$
(B part), $s^{i}\phi_{i}$ (S part) and $\left(m^{2}\right)_{j}^{i}\phi_{i}\phi_{j}^{*}$
(M2 part). 
\end{itemize}
\texttt{ShowLagrangian} is completely independent of \texttt{BetaFunctions1L}
and \texttt{BetaFunctions2L}, so it can be called at any time to generate
the superpotential and soft-SUSY-breaking Lagrangian of a given model.

Finally, notice that in order to build $\mathscr{L}$ it is necessary
to assume a particular basis for the representations/fields of the
model. Quantities such as the $\beta$-functions clearly do not depend
on this choice. However, it is important to know that \texttt{Susyno}
can use unconventional basis for the representations. The following
is a good example. According to the program, two fields, A=\{a{[}1{]},
a{[}2{]}, a{[}3{]}\} and B=\{b{[}1{]}, b{[}2{]}, b{[}3{]}\}, transforming
as $\boldsymbol{3}$ and $\overline{\boldsymbol{3}}$ of SU(3) will
form the invariant a{[}1{]}b{[}3{]} - a{[}2{]} b{[}2{]} + a{[}3{]}b{[}1{]},
not a{[}1{]}b{[}1{]} + a{[}2{]} b{[}2{]} + a{[}3{]}b{[}3{]}. Both
are valid expressions as long as they are used in a consistent way
- it is just a matter of basis choice. Thus, when analysing the output
of \texttt{ShowLagrangian} we must distinguish between irrelevant
basis choices and relevant normalisation conventions for the parameters.
The $\beta$-functions are only sensitive to the latter ones.

\section{\label{sec:Tests-made}Tests/validation of \texttt{Susyno}}

The output of \texttt{Susyno} was confronted with the analysis of some models
 available in the literature. In particular, the RGEs generated by \texttt{Susyno}
 were compared with the results for the MSSM \cite{MartinVaughn:RGEs}, the
 R-parity violating MSSM \cite{Dreiner:1999}, the general NMSSM
\cite{Ellwanger:2009dp} and some SU(5)-based models \cite{Borzumati:2009hu}. 

In general the program's RGEs are consistent with the results
collected in the above publications, but in some cases
differences were found. These have been collected in the program's
webpage.

\section*{Acknowledgments}

The author would like to thank Jorge Rom\~ao and Ana M. Teixeira for the
many suggestions and encouragement given.
This work was supported by the \textit{Funda\c c\~ao para a Ci\^encia
e a Tecnologia} under the grant SFRH/BD/47795/2008.

\appendix

\setcounter{table}{0}
\section{List of available functions\label{sec:List-of-available}}

As in most programs, the code \texttt{Susyno} is spread over many internal
functions. Due to their nature, some of these functions may be useful
on their own, and they were thus built in a user-friendly way, and are documented.

Below is the list of functions that can be called
directly by the user in Mathematica's front-end, 
followed by a brief description. The built-in
help system describes in detail how to use them.
\begin{itemize}
\item \texttt{Adjoint} - Computes the Dynkin coefficients of the adjoint
representation of a group;
\item \texttt{BetaFunctions1L} - Computes the 1-loop $\beta$-functions
of a SUSY model;
\item \texttt{BetaFunctions2L} - Computes the 2-loop $\beta$-functions
of a SUSY model;
\item \texttt{CanonicalForm} - Simplifies an expression written in Einstein's
notation;
\item \texttt{Casimir} - Computes the quadratic Casimir of a representation;
\item \texttt{CartanMatrix} - Computes the Cartan matrix of any simple group;
\item \texttt{DimR} - Computes the dimension of a representation;
\item \texttt{Invariants} - Computes (in some basis) the invariant combination(s)
of a product of one, two or three representations;
\item \texttt{ListContract} - Efficiently calculates traces of multi-index
sparse tensors;
\item \texttt{PositiveRoots} - Computes the positive roots of a group;
\item \texttt{ReduceRepProduct} - Decomposes a direct product representation
in its irreducible parts~\cite{DMSnow};
\item \texttt{RepMatrices} - Computes (in some basis) the explicit representation
matrices;
\item \texttt{ShowLagrangian} - Generates the Lagrangian ($W$ and $\mathscr{L}_{soft}$)
of a SUSY model;
\item \texttt{Weights} - Computes the weights of a representation, including
degeneracy.
\end{itemize}

\section{Parameters of the MSSM \label{sec:Parameters-of-the-MSSM}}

For completeness, we include here the superpotential and soft-SUSY-breaking Lagrangian of the MSSM, following the conventions
of~\cite{MartinVaughn:RGEs}.
\begin{eqnarray}
W & = & \hat uY_{u}\hat Q\cdot \hat H_{u}+
\hat d Y_{d}\hat Q\cdot \hat H_{d}+\hat eY_{e}\hat L\cdot \hat H_{d}+
\mu \hat H_{u}\cdot \hat H_{d} \label{eq:app:W}\\
-\mathscr{L}_{soft} & = & \left(\tilde{u}h_{u}\tilde{Q}\cdot{H}_{u}+
\tilde{d}h_{d}\tilde{Q}\cdot{H}_{d}+\tilde{e}h_{e}\tilde{L}\cdot{H}_{d}+
B {H}_{u}\cdot{H}_{d}+\textrm{h.c.}\right)\nonumber \\
 &  &
+\tilde{Q}^{\dagger}m_{Q}^{2}\tilde{Q}+\tilde{L}^{\dagger}m_{L}^{2}\tilde{L}+
\tilde{u}m_{u}^{2}\tilde{u}^{\dagger}+\tilde{d}m_{d}^{2}\tilde{d}^{\dagger}+
\tilde{e}m_{e}^{2}\tilde{e}^{\dagger}+m_{H_{u}}^{2}{H}_{u}^{\dagger}{H}_{u}+
m_{H_{d}}^{2} {H}_{d}^{\dagger}{H}_{d}\nonumber \\
 &  &
+\left(\frac{1}{2}M_{a}\lambda_{a}\lambda_{a}+\textrm{h.c.}\right)
\label{eq:app:Lsoft}
\end{eqnarray}
In the above, and as usual, hats denote superfields and tildes 
the scalar component of each chiral superfield. The contraction
of two SU(2) doublets (denoted by $\cdot$) follows the convention
$H_{u}\cdot H_{d}=H_{u}^{+}H_{d}^{-}-H_{u}^{0}H_{d}^{0}$. In addition
to the parameters entering $W$ and $\mathscr{L}_{soft}$, one still has
the gauge couplings $g_{a}$. 
Table \ref{tab:ParametersMSSM} summarises the matching of the physical
notation of Eqs.~(\ref{eq:app:W},\ref{eq:app:Lsoft}) 
with the one used by \texttt{Susyno}.

\begin{table}
\begin{centering}
\begin{tabular}{cc}
Parameter & \texttt{Susyno}'s notation\tabularnewline
\hline
$g_{1}$, $g_{2}$, $g_{3}$ & \texttt{g{[}1{]}, g{[}2{]}, g{[}3{]}}\tabularnewline
$M_{1}$, $M_{2}$, $M_{3}$ & \texttt{M{[}1{]}, M{[}2{]}, M{[}3{]}}\tabularnewline
$\left(Y_{u}\right)_{ij}$ & \texttt{y{[}1,2,6,1,j,i{]}}\tabularnewline
$\left(Y_{d}\right)_{ij}$ & \texttt{y{[}1,3,7,1,j,i{]}}\tabularnewline
$\left(Y_{e}\right)_{ij}$ & \texttt{y{[}4,5,7,1,j,i{]}}\tabularnewline
$\mu$ & \texttt{mu{[}6,7,1{]}}\tabularnewline
$\left(h_{u}\right)_{ij}$ & \texttt{h{[}1,2,6,1,j,i{]}}\tabularnewline
$\left(h_{d}\right)_{ij}$ & \texttt{h{[}1,3,7,1,j,i{]}}\tabularnewline
$\left(h_{e}\right)_{ij}$ & \texttt{h{[}4,5,7,1,j,i{]}}\tabularnewline
$B$ & \texttt{b{[}6,7,1{]}}\tabularnewline
$\left(m_{Q}^{2}\right)_{ij}$ & \texttt{m2{[}1,1,1,j,i{]}}\tabularnewline
$\left(m_{u}^{2}\right)_{ij}$ & \texttt{m2{[}2,2,1,i,j{]}}\tabularnewline
$\left(m_{d}^{2}\right)_{ij}$ & \texttt{m2{[}3,3,1,i,j{]}}\tabularnewline
$\left(m_{L}^{2}\right)_{ij}$ & \texttt{m2{[}4,4,1,j,i{]}}\tabularnewline
$\left(m_{e}^{2}\right)_{ij}$ & \texttt{m2{[}5,5,1,i,j{]}}\tabularnewline
$m_{H_{u}}^{2}$ & \texttt{m2{[}6,6,1{]}}\tabularnewline
$m_{H_{d}}^{2}$ & \texttt{m2{[}7,7,1{]}}\tabularnewline
\end{tabular}
\par\end{centering}

\caption{\label{tab:ParametersMSSM}Parameters of the MSSM assuming the field
ordering \{Q,u,d,L,e,Hu,Hd\} and the group ordering
\{U1,SU2,SU3\}.}

\end{table}

\section{More on groups and representations\label{sec:More-on-groups-reps}}

A simple complex Lie algebra is specified by a Cartan matrix~\cite{RobertCahn}
which is a square matrix with diagonal entries all equal to 2 and
off-diagonal entries equal to 0, -1, -2 or -3. For example, the Cartan
matrix of SO(10) is
\begin{equation}
\left(\begin{array}{ccccc}
2 & -1 & 0 & 0 & 0\\
-1 & 2 & -1 & 0 & 0\\
0 & -1 & 2 & -1 & -1\\
0 & 0 & -1 & 2 & 0\\
0 & 0 & -1 & 0 & 2\end{array}\right)
\end{equation}
The Cartan matrix is the ingredient required by \texttt{Susyno}
whenever a group is being asked as input (the exception is U(1), which is 
taken to be just \texttt{U1} or \texttt{\{\}}). Although the
user can manually provide the Cartan matrices, the program already
has a built-in function, \texttt{CartanMatrix}, that can compute 
them for any simple group. The syntax is\\

\noindent\texttt{CartanMatrix{[}{}"group name",familyIndex{]}}~\\
\texttt{}~\\
where \texttt{{}"group name"} can be \texttt{{}"SO"},
\texttt{{}"SU"}, \texttt{{}"SP"}, \texttt{{}"G"}, \texttt{{}"F"}
or \texttt{{}"E"} and \texttt{familyIndex} should be an integer.
For example\\
\\
\noindent\texttt{CartanMatrix{[}{}"SO",10{]}}\\
\\
will return the above $5\times5$ Cartan matrix. This function
will work for virtually any simple group. However, this is still a
heavy syntax, so the program assigns to some variables (\texttt{SU2},
\texttt{SU3}, \texttt{SU5}, \texttt{SO10}, etc.) the corresponding
Cartan matrices (e.g., \texttt{SO10} is the same as \texttt{CartanMatrix{[}{}"SO",10{]}}).
For the case of SO($n$), SU($n$) with $n>32$ and Sp($2m$) with $m>16$ no
variables were set, and the \texttt{CartanMatrix} function should be used.\\

As mentioned in the text, \texttt{Susyno} requires that each field be
specified as a list
of hypercharges and representations under the simple gauge factor
groups. We will focus now on this last element: representations of
simple groups.

A representation of a (simple) group is often labeled by its dimension
(e.g., $\boldsymbol{3}$ of SU(3), $\boldsymbol{54}$ of SO(10), ...)
but this can be ambiguous. To avoid the problem, we should use Dynkin
coefficients which are a list of $n$ non-negative integers. This
number $n$ is known as the group rank and is equal to the size of
the (square) Cartan matrix of the group. For instance, since the Cartan
matrix of SO(10) is $5\times5$, all representations of the group
are of the form\\
\\
 \texttt{\{i1,i2,i3,i4,i5\}}\\
\\
(\texttt{i1},..., \texttt{i5} are non-negative integers).

There are functions in \texttt{Susyno} that compute properties of
the representations (e.g., \texttt{DimR} calculates the dimension
of a representation, \texttt{ReduceRepProduct} reduces products of
representations) and they are documented in the built-in help files
(see also \ref{sec:List-of-available}).
These should be enough to allow recognising a representation by its
Dynkin coefficients; however, should the user wish to consult lists of
representations, these are available in the literature~\cite{Slansky:1981}.
Some of the most frequently used representations were shown in 
table~\ref{tab:Table_representations}.

\section{Renormalisation group equations of a general model
\label{sec:General-renormalisation-group}}

Susyno uses the generic two loop RGEs of \cite{MartinVaughn:RGEs,Yamada:1994id}.
For completeness these results are reproduced here (for a model based
on a simple gauge group $G$). \\
In general a parameter $X$ evolves according to the equation
\begin{eqnarray}
\frac{d}{dt}X & = & \frac{1}{16\pi^{2}}\beta_{X}^{\left(1\right)}+
\frac{1}{\left(16\pi^{2}\right)^{2}}\beta_{X}^{\left(2\right)}
\end{eqnarray}
where $t=\log Q$ ($Q$ being the renormalisation scale parameter).
The $\beta$-functions for the several parameters (gauge coupling
constants, superpotential parameters and soft-SUSY-breaking
parameters) are as follows.

\subsection*{Gauge coupling constants}

\begin{eqnarray}
\beta_{g}^{\left(1\right)} & = & g^{3}\left[S\left(R\right)-3C\left(G\right)\right]\\
\beta_{g}^{\left(2\right)} & = & g^{5}\left\{
-6\left[C\left(G\right)\right]^{2}+2C\left(G\right)S\left(R\right)+4S\left(R\right)C\left(R\right)\right\}
-g^{3}\frac{C\left(k\right)}{d\left(G\right)}Y^{ijk}Y_{ijk}
\end{eqnarray}

\subsection*{Superpotential parameters}

\begin{eqnarray}
\left[\beta_{Y}^{\left(1\right)}\right]^{ijk} & = &
Y^{ijp}\gamma_{p}^{\left(1\right)k}+\left(k\leftrightarrow
i\right)+\left(k\leftrightarrow j\right)\\ 
\left[\beta_{Y}^{\left(2\right)}\right]^{ijk} & = &
Y^{ijp}\gamma_{p}^{\left(2\right)k}+\left(k\leftrightarrow
i\right)+\left(k\leftrightarrow j\right)\\ 
\nonumber \\\left[\beta_{\mu}^{\left(1\right)}\right]^{ij} & = &
\mu^{ip}\gamma_{p}^{\left(1\right)j}+\left(j\leftrightarrow
i\right)\\ 
\left[\beta_{\mu}^{\left(2\right)}\right]^{ij} & = &
\mu^{ip}\gamma_{p}^{\left(2\right)j}+\left(j\leftrightarrow
i\right)\\ 
\nonumber \\\left[\beta_{L}^{\left(1\right)}\right]^{i} & = & L^{p}\gamma_{p}^{\left(1\right)i}\\
\left[\beta_{L}^{\left(2\right)}\right]^{i} & = &
L^{p}\gamma_{p}^{\left(2\right)i}
\end{eqnarray}
\noindent where
\begin{eqnarray}
\gamma_{i}^{\left(1\right)j} & = & \frac{1}{2}Y_{ipq}Y^{jpq}-2g^{2}\delta_{i}^{j}C\left(i\right)\\
\gamma_{i}^{\left(2\right)j} & = &
-\frac{1}{2}Y_{imn}Y^{npq}Y_{pqr}Y^{mrj}+g^{2}Y_{ipq}Y^{jpq}\left[2C\left(p\right)-C\left(i\right)\right]\nonumber
\\ 
 &  &
+2\delta_{i}^{j}g^{4}\left[C\left(i\right)S\left(R\right)+
2C\left(i\right)^{2}-3C\left(G\right)C\left(i\right)\right] 
\end{eqnarray}

\subsection*{Soft-SUSY-breaking Lagrangian parameters}

\begin{eqnarray}
\beta_{M}^{\left(1\right)} & = & g^{2}\left[2S\left(R\right)-6C\left(G\right)\right]M\\
\beta_{M}^{\left(2\right)} & = &
g^{4}\left[-24C\left(G\right)^{2}+8C\left(G\right)S\left(R\right)+
16S\left(R\right)C\left(R\right)\right]M+2g^{2}\frac{C\left(k\right)}{d\left(G\right)}
\left(h^{ijk}-MY^{ijk}\right)Y_{ijk} \nonumber
\\ && 
\end{eqnarray}
\begin{eqnarray}
\left[\beta_{h}^{\left(1\right)}\right]^{ijk} & = &
\frac{1}{2}h^{ijl}Y_{lmn}Y^{mnk}+Y^{ijl}Y_{lmn}h^{mnk}-2\left(h^{ijk}-2MY^{ijk}\right)g^{2}C(k)+(k\leftrightarrow
i)+(k\leftrightarrow j)\nonumber \\
 && \\
\left[\beta_{h}^{\left(2\right)}\right]^{ijk} & = &
-\frac{1}{2}h^{ijl}Y_{lmn}Y^{npq}Y_{pqr}Y^{mrk}-Y^{ijl}Y_{lmn}Y^{npq}Y_{pqr}h^{mrk}-
Y^{ijl}Y_{lmn}h^{npq}Y_{pqr}Y^{mrk}\nonumber
\\ 
 &  &
+\left(h^{ijl}Y_{lpq}Y^{pqk}+2Y^{ijl}Y_{lpq}h^{pqk}-2MY^{ijl}Y_{lpq}Y^{pqk}\right)g^{2}
\left[2C(p)-C(k)\right]\nonumber
\\ 
 &  &
+\left(2h^{ijk}-8MY^{ijk}\right)g^{4}\left[C(k)S(R)+2C(k)^{2}-3C(G)C(k)\right]+
(k\leftrightarrow i)+(k\leftrightarrow j)
\nonumber
\\ && 
\end{eqnarray}
\begin{eqnarray}
\left[\beta_{b}^{\left(1\right)}\right]^{ij} & = &
\frac{1}{2}b^{il}Y_{lmn}Y^{mnj}+
\frac{1}{2}Y^{ijl}Y_{lmn}b^{mn}+\mu^{il}Y_{lmn}h^{mnj}-2\left(b^{ij}-
2M\mu^{ij}\right)g^{2}C(i)+(i\leftrightarrow j)\nonumber \\
&&\\
\left[\beta_{b}^{\left(2\right)}\right]^{ij} & = &
-\frac{1}{2}b^{il}Y_{lmn}Y^{pqn}Y_{pqr}Y^{mrj}-
\frac{1}{2}Y^{ijl}Y_{lmn}b^{mr}Y_{pqr}Y^{pqn}-\frac{1}{2}Y^{ijl}Y_{lmn}\mu^{mr}Y_{pqr}h^{pqn}
\nonumber \\
 &  &
-\mu^{il}Y_{lmn}h^{npq}Y_{pqr}Y^{mrj}-\mu^{il}Y_{lmn}Y^{npq}Y_{pqr}h^{mrj}+2Y^{ijl}Y_{lpq}
\left(b^{pq}-\mu^{pq}M\right)g^{2}C(p)\nonumber \\
 &  & +\left(b^{il}Y_{lpq}Y^{pqj}+2\mu^{il}Y_{lpq}h^{pqj}-
2\mu^{il}Y_{lpq}Y^{pqj}M\right)g^{2}\left[2C(p)-C(i)\right]\nonumber \\
 &  &
+\left(2b^{ij}-8\mu^{ij}M\right)g^{4}\left[C(i)S(R)+2C(i)^{2}-3C(G)C(i)\right]+(i\leftrightarrow
j)
\end{eqnarray}
\begin{eqnarray}
\left[\beta_{s}^{\left(1\right)}\right]^{i} & = &
\frac{1}{2}Y^{ipq}Y_{pqr}s^{r}+h^{ipq}Y_{pqr}L^{r}+
\mu^{ir}Y_{rpq}b^{pq}+2Y^{imn}(m^{2})_{m}^{l}\mu_{nl}+h^{ipq}b_{pq}\\
\left[\beta_{s}^{\left(2\right)}\right]^{i} & = &
2g^{2}C\left(q\right)Y^{ipq}Y_{pqr}s^{r}-
\frac{1}{2}Y^{irn}Y_{npq}Y^{pqm}Y_{mrl}s^{l}-4g^{2}C\left(q\right)\left(Y^{ipq}M-h^{ipq}\right)
Y_{pql}L^{l}\nonumber \\
 &  &
-\left(Y^{imn}Y_{npq}h^{pqr}Y_{rml}+h^{imn}Y_{npq}Y^{pqr}Y_{rml}\right)L^{l}
-4g^{2}C\left(q\right)\mu^{il}Y_{lpq}\left(\mu^{pq}M-b^{pq}\right)\nonumber \\
 &  &
-\mu^{il}\left(Y_{lmn}h^{npq}Y_{pqr}\mu^{rm}+Y_{lmn}Y^{npq}Y_{pqr}b^{rm}\right)
+4g^{2}C\left(q\right)\left[2Y^{ipq}\mu_{pq}|M|^{2}-Y^{ipq}b_{pq}M\right.\nonumber \\
 &  &
  \left.-h^{ipq}\mu_{pq}M^{*}+h^{ipq}b_{pq}+Y^{ipq}(m^{2})_{p}^{r}\mu_{rq}+
Y^{ipr}(m^{2})_{r}^{q}\mu_{pq}\right]-Y^{imn}Y_{npq}h^{pql}b_{lm}\nonumber \\
 &  &
-h^{imn}Y_{npq}Y^{pql}b_{lm}-Y^{imn}h_{npq}h^{pql}\mu_{lm}-h^{imn}h_{npq}Y^{pql}\mu_{lm}-
Y^{imn}(m^{2})_{m}^{l}\mu_{lr}Y^{rpq}Y_{pqn}\nonumber \\
 &  &
-Y^{imn}Y_{npq}Y^{pql}(m^{2})_{l}^{r}\mu_{rm}-Y^{imn}(m^{2})_{n}^{l}Y_{lpq}Y^{pqr}\mu_{rm}-
2Y^{imn}Y_{npq}(m^{2})_{r}^{q}Y^{rpl}\mu_{lm}
\nonumber
\\ && 
\end{eqnarray}
\begin{eqnarray}
\left[\beta_{m^{2}}^{\left(1\right)}\right]_{i}^{j} & = & 
\frac{1}{2}Y_{ipq}Y^{pqn}{(m^{2})}_{n}^{j}+\frac{1}{2}Y^{jpq}Y_{pqn}{(m^{2})}_{i}^{n}
+2Y_{ipq}Y^{jpr}{(m^{2})}_{r}^{q}+h_{ipq}h^{jpq}\nonumber \\
 &  & -8\delta_{i}^{j}\left|M\right|^{2}g^{2}C(i)+
2g^{2}\boldsymbol{t}_{i}^{\boldsymbol{A}j}\textrm{Tr}\left(\boldsymbol{{\bf t}^{A}}m^{2}\right)\\
\left[\beta_{m^{2}}^{\left(2\right)}\right]_{i}^{j} & = & 
-\frac{1}{2}{(m^{2})}_{i}^{l}Y_{lmn}Y^{mrj}Y_{pqr}Y^{pqn}-
\frac{1}{2}{(m^{2})}_{l}^{j}Y^{lmn}Y_{mri}Y^{pqr}Y_{pqn}-Y_{ilm}Y^{jnm}{(m^{2})}_{r}^{l}Y_{npq}Y^{rpq}\nonumber \\
 &  &
-Y_{ilm}Y^{jnm}{(m^{2})}_{n}^{r}Y_{rpq}Y^{lpq}-Y_{ilm}Y^{jnr}{(m^{2})}_{n}^{l}Y_{pqr}Y^{pqm}-
2Y_{ilm}Y^{jln}Y_{npq}Y^{mpr}{(m^{2})}_{r}^{q}\nonumber \\
 &  &
-Y_{ilm}Y^{jln}h_{npq}h^{mpq}-h_{ilm}h^{jln}Y_{npq}Y^{mpq}-h_{ilm}Y^{jln}Y_{npq}h^{mpq}
-Y_{ilm}h^{jln}h_{npq}Y^{mpq}\nonumber \\
 &  &
+\biggl[{(m^{2})}_{i}^{l}Y_{lpq}Y^{jpq}+Y_{ipq}Y^{lpq}{(m^{2})}_{l}^{j}+
4Y_{ipq}Y^{jpl}{(m^{2})}_{l}^{q}+2h_{ipq}h^{jpq}-2h_{ipq}Y^{jpq}M\nonumber \\
 &  &
-2Y_{ipq}h^{jpq}M^{*}+4Y_{ipq}Y^{jpq}\left|M\right|^{2}\biggr]g^{2}
\left[C(p)+C(q)-C(i)\right]-2g^{2}\boldsymbol{t}_{i}^{\boldsymbol{A}j}
(\boldsymbol{{\bf t}^{A}}m^{2})_{r}^{l}Y_{lpq}Y^{rpq}\nonumber \\
 &  & +8g^{4}{\bf t}_{i}^{Aj}\textrm{Tr}\left[\boldsymbol{{\bf
      t}^{A}}C(r)m^{2}\right]
+\delta_{i}^{j}g^{4}\left|M\right|^{2}\left[24C(i)S(R)+48C(i)^{2}-72C(G)C(i)\right]\nonumber \\
 &  & +8\delta_{i}^{j}g^{4}C(i)\left\{
\textrm{Tr}\left[S(r)m^{2}\right]
-C(G)\left|M\right|^{2}\right\} \end{eqnarray}
\\
A few comments and clarifications are still in order concerning 
the variables appearing in the different $\beta$-functions:
\begin{itemize}
\item $Y_{ijk}=\left(Y^{ijk}\right)^{*}$, $h_{ijk}=\left(h^{ijk}\right)^{*}$,
$\mu_{ij}=\left(\mu^{ij}\right)^{*}$ and $b_{ij}=\left(b^{ij}\right)^{*}$;
\item $d\left(G\right)$ = Dimension of the adjoint representation of group
$G$;
\item $C\left(i\right)$ = Quadratic Casimir invariant of the representation
of the chiral superfield with index $i$;
\item $C\left(G\right)$ = Quadratic Casimir invariant of the adjoint representation
of group $G$;
\item $S\left(R\right)$ = Dynkin index summed over all chiral multiplets.
However $S\left(R\right)C\left(R\right)$ should be interpreted as
the sum of Dynkin indices weighted by the quadratic Casimir invariant;
\item $\boldsymbol{t^{A}}$ = Representation matrices under the gauge group
$G$. Terms with $\boldsymbol{t^{A}}$ are only relevant for U(1)
groups;
\item In $\beta_{m^{2}}^{\left(1\right)}$ and $\beta_{m^{2}}^{\left(2\right)}$,
the traces should be understood as traces over all chiral superfields.
\end{itemize}
If the gauge group is not simple but rather a direct product of simple
groups and U(1) factors, the equations shown here must be adapted.
The modifications are fairly straightforward, and can be found in~\cite{MartinVaughn:RGEs}.

\end{document}